\documentclass[onecolumn, 11pt]{IEEEtran}
\usepackage{graphicx}
\usepackage{cite}
\usepackage{amsmath}
\usepackage{subfigure}
\usepackage{setspace}
\IEEEoverridecommandlockouts

\ifCLASSINFOpdf
\else
\fi

\begin{document}
\title{Relay-Assisted Multiple Access with Multi-Packet Reception Capability and Simultaneous Transmission and Reception}

\author{\IEEEauthorblockN{Nikolaos Pappas, Anthony Ephremides, Apostolos Traganitis\thanks{N. Pappas and A. Traganitis are with the Computer Science Department, University of Crete Greece and Institute of Computer Science, Foundation for Research and Technology - Hellas (FORTH). N. Pappas at that time was visiting Department of Electrical and Computer Engineering and Institute for Systems Research University of Maryland.}
\thanks{A. Ephremides is with the Department of Electrical and Computer Engineering and Institute for Systems Research University of Maryland, College Park, MD 20742.}
\thanks{N. Pappas was supported by "HRAKLEITOS II - University of Crete", NSRF (ESPA) (2007-2013) and is co-funded by the European Union and national resources. This work was supported in part by MURI grant W911NF-08-1-0238, NSF grant CCF-0728966, and ONR grant N000141110127.}}
\IEEEauthorblockA{\\E-mail: npapas@ics.forth.gr, etony@umd.edu, tragani@ics.forth.gr}
}

\maketitle

\begin{abstract}
In this work we examine the operation of a node relaying packets from a number of users to a destination node. We assume multi-packet reception capabilities for the relay and the destination node. The relay node can transmit and receive at the same time, so the problem of self interference arises. The relay does not have packets of its own and the traffic at the source nodes is considered saturated. The relay node stores a source packet that it receives successfully in its queue when the transmission to the destination node has failed. We obtain analytical expressions for the characteristics of the relay's queue (such as arrival and service rate of the relay's queue), the stability condition and the average length of the queue as functions of the probabilities of transmissions, the self interference coefficient and the outage probabilities of the links. We study the impact of the relay node and the self interference coefficient on the throughput per user-source as well as the aggregate throughput.
\end{abstract}

\section{Introduction}
\label{sec:intro}
The classical relay channel was originally introduced by van der Meulen~\cite{b:Muelen}. The first works on the relay channel were based on information theoretical formulations as in~\cite{b:CoverGamal} and~\cite{b:Sadek}. Recently several works investigated relaying capability at the MAC layer~\cite{b:Sadek},~\cite{b:Simeone},~\cite{b:Rong1},~\cite{b:Rong2}. The classical analysis of random multiple access schemes like slotted ALOHA~\cite{b:Bertsekas} has focused on the so called collision model. Random access with multi-packet reception (MPR) has attracted attention recently~\cite{b:AlohaVerdu},~\cite{b:Angel},~\cite{b:Naware}. All these previous approaches come together in the model that we consider.

In wireless networks when a node transmits and receives simultaneously the problem of self interference arises. Information theoretic aspects of this problem can be found at the work of Shannon on~\cite{b:Shannon2way}, although the capacity region of the two-way channel is not known for the general case~\cite{b:Cover}. There are some techniques that allow the possibility of perfect self interference cancelation~\cite{b:Cover}. In practice though, there are technological limitations~\cite{b:selflimit1}-~\cite{b:selflimit2} which can limit the accuracy of the self interference cancelation. Various methods for performing self interference cancelation at the nodes' receivers can be found in~\cite{b:selfcancel1} and~\cite{b:selfcancel2}. The conclusion is that there is a trade off between transceiver complexity and the accuracy of the self interference cancelation.

In this work we examine the operation of a node relaying packets from a number of users-sources to a destination node as shown in Fig.~\ref{fig:netmodel}, and is an extension of~\cite{b:Pappas} and~\cite{b:Pappas-WiOpt}. We assume MPR capability for the relay and the destination node. The relay node can transmit and receive at the same time.
We assume random access to the channel, time is considered slotted, and each packet transmission takes one time slot. The wireless channel between the nodes in the network is modeled by a Rayleigh narrowband flat-fading channel with additive Gaussian noise. A user's transmission is successful if the received signal to interference plus noise ratio ($SINR$) is above a threshold $\gamma$. We also assume that acknowledgements (ACKs) are instantaneous and error free. The relay does not have packets of its own and the sources are considered saturated with unlimited amount of traffic. We do not consider any specific self interference cancelation mechanism, because it is out of the scope of this work. The self interference cancelation at the relay is modeled as a variable power gain.

We obtain analytical expressions for the characteristics of the relay's queue (such as arrival and service rates), we study the stability condition and the average length of the queue as functions of the probabilities of transmission, the self interference coefficient and the outage probabilities of the links. We study the impact of the relay node and the self interference coefficient on the throughput per user-source and the aggregate throughput.

Section~\ref{sec:sysmod} describes the system model, in Section~\ref{sec:analysis} we study the characteristics of the relay's queue and we derive the equations for the throughput per user and the aggregate throughout. We present the numerical results in Section~\ref{sec:results} and, finally, our conclusions are given in Section~\ref{sec:conclusions}.

\section{System Model}
\label{sec:sysmod}
\subsection{Network Model}
\begin{center}
\begin{figure}[ht]
\begin{minipage}[b]{0.3\linewidth}
\centering
\includegraphics[scale=0.45]{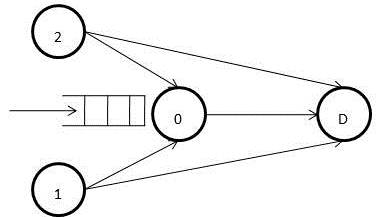}
\caption{The simple network model}
\label{fig:netmodel}
\end{minipage}
\begin{minipage}[b]{0.3\linewidth}
\centering
\includegraphics[scale=0.4]{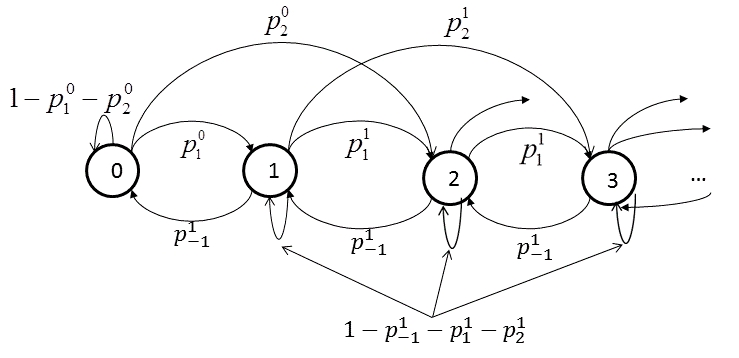}
\caption{Markov Chain model for the two-user case}
\label{fig:mc_rxtx}
\end{minipage}
\end{figure}
\end{center}

We consider a network with $N$ sources, one relay node and a single destination node. The sources transmit packets to the destination with the cooperation of the relay; the case of $N=2$ is depicted in Fig.~\ref{fig:netmodel}. We assume that the queues of the two sources are saturated (i.e. there are no external arrivals but unlimited packet volume in the buffers); the relay does not have packets of its own, and just forwards the packets that it has received from the two users. The relay node stores a source packet that it receives successfully in its queue when the direct transmission to the destination node has failed. We assume random access to the channel. Each of the receivers (relay and destination) is equipped with multiuser detectors, so that they may decode packets successfully from more than one transmitter at a time. The relay node can receive and transmit packets simultaneously.

%

\subsection{Physical Layer Model}
The MPR channel model used in this paper is a generalized form of the packet erasure model. We assume that a packet transmitted by $i$ is successfully received by $j$ if and only if $SINR(i,j)\geq \gamma_{j}$, where $\gamma_{j}$ is a threshold characteristic of node $j$. The wireless channel is subject to fading; let $P_{tx}(i)$ be the transmitting power at node $i$ and $r(i,j)$ be the distance between $i$ and $j$. The received power by $j$ when $i$ transmits is $P_{rx}(i,j)=A(i,j)g(i,j)$ where $A(i,j)$ is a random variable representing channel fading. Under Rayleigh fading, it is known~\cite{b:Tse} that $A(i,j)$ is exponentially distributed. The received power factor $g(i,j)$ is given by $g(i,j)=P_{tx}(i)(r(i,j))^{-\alpha}$ where $\alpha$ is the path loss exponent with typical values between $2$ and $4$. We model the self interference by a scalar $g \in [0,1]$. We refer to the $g$ as the self interference coefficient. When $g=1$, no self interference cancelation technique is used and $g=0$ when there is perfect self interference cancelation.
The success probability in the link $ij$ is given by:

\begin{equation}
\label{eq:succprob}
P_{i/T}^{j}=\exp\left(-\frac{\gamma_{j}\eta_{j}}{v(i,j)g(i,j)}\right) \left(1+\gamma_{j}(r(i,j))^{\alpha}g \right)^{-m} \prod_{k\in T\backslash \left\{i,j\right\}}{\left(1+\gamma_{j}\frac{v(k,j)g(k,j)}{v(i,j)g(i,j)}\right)}^{-1}
\end{equation}

where T is the set of transmitting nodes at the same time, $v(i,j)$ is the parameter of the Rayleigh random variable for fading; $m=1$ when $j \in T$ and $m=0$ else. The analytical derivation for this success probability can be found in~\cite{b:Nguyen} and~\cite{b:Codreanu}.

\section{Analysis}
\label{sec:analysis}
In this section we derive the equations for the characteristics of the relay's queue, such as the arrival and service rates, the stability conditions, and the average queue length. We will provide an analysis for two cases: first, when the network consists of two users (non-symmetric) and the second is for $n>2$ symmetric users.

\subsection{Two-user case}

\subsubsection{Computation of the average arrival and service rate}
The service rate is given by:
\begin{equation}
\label{eq:m2}
\begin{aligned}
\mu=q_{0}(1-q_{1})(1-q_{2})P_{0/0}^{d}+q_{0}q_{1}(1-q_{2})P_{0/0,1}^{d}+q_{0}q_{2}(1-q_{1})P_{0/0,2}^{d}+q_{0}q_{1}q_{2}P_{0/0,1,2}^{d}
\end{aligned}
\end{equation}
where $q_{0}$ is the transmission probability of the relay given that it has packets in its queue, $q_{i}$ for $i \neq 0$ is the transmission probability for the $i$-th user. The term $P_{i/i,k}^{j}$ is the success probability of link $ij$ when the transmitting nodes are $i$ and $k$ and can be calculated based on (\ref{eq:succprob}).

The average arrival rate $\lambda$ of the queue is given by:
\begin{equation}
\label{eq:lamda2}
\lambda=P\left(Q=0\right)\lambda_{0}+P\left(Q>0\right)\lambda_{1}
\end{equation}

Where $\lambda_{0}$ is the average arrival rate at the relay's queue when the queue is empty and $\lambda_{1}$ when it's not.
$\lambda_{0}=r_{1}^{0}+2r_{2}^{0}$, where $r_{i}^{0}$ is the probability of receiving $i$ packets given that the queue is empty. The expressions for $r_{i}^{0}$ are given by:

{\scriptsize
\begin{equation}
\label{eq:r10}
\begin{aligned}
r_{1}^{0}=q_{1}(1-q_{2})(1-P_{1/1}^{d})P_{1/1}^{0}+q_{2}(1-q_{1})(1-P_{2/2}^{d})P_{2/2}^{0}+q_{1}q_{2}(1-P_{1/1,2}^{d})P_{1/1,2}^{0}P_{2/1,2}^{d}+\\
+q_{1}q_{2}(1-P_{2/1,2}^{d})P_{2/1,2}^{0}P_{1/1,2}^{d}+q_{1}q_{2}(1-P_{1/1,2}^{d})P_{1/1,2}^{0}(1-P_{2/1,2}^{d})(1-P_{2/1,2}^{0})+q_{1}q_{2}(1-P_{2/1,2}^{d})P_{2/1,2}^{0}(1-P_{1/1,2}^{d})(1-P_{1/1,2}^{0})
\end{aligned}
\end{equation}
}
\begin{equation}
\label{eq:r20}
\begin{aligned}
r_{2}^{0}=q_{1}q_{2}(1-P_{1/1,2}^{d})(1-P_{2/1,2}^{d})P_{1/1,2}^{0}P_{2/1,2}^{0}
\end{aligned}
\end{equation}

Accordingly, $\lambda_{1}=r_{1}^{1}+2r_{2}^{1}$, where $r_{i}^{1}$ is the probability of receiving $i$ packets when the queue is not empty.
The expressions for the $r_{i}^{1}$ are lengthy and given by:

{\scriptsize
\begin{equation}
\label{eq:r11}
\begin{aligned}
r_{1}^{1}=(1-q_{0})q_{1}(1-q_{2})(1-P_{1/1}^{d})P_{1/1}^{0}+q_{0}q_{1}(1-q_{2})(1-P_{1/0,1}^{d})P_{1/0,1}^{0}+(1-q_{0})q_{2}(1-q_{1})(1-P_{2/2}^{d})P_{2/2}^{0}+q_{0}q_{2}(1-q_{1})(1-P_{2/0,2}^{d})P_{2/0,2}^{0}+\\
+(1-q_{0})q_{1}q_{2}(1-P_{1/1,2}^{d})P_{1/1,2}^{0}(1-P_{2/1,2}^{d})(1-P_{2/1,2}^{0})+q_{0}q_{1}q_{2}(1-P_{1/0,1,2}^{d})P_{1/0,1,2}^{0}(1-P_{2/0,1,2}^{d})(1-P_{2/0,1,2}^{0})+\\
+(1-q_{0})q_{1}q_{2}(1-P_{1/1,2}^{d})P_{1/1,2}^{0}P_{2/1,2}^{d}+q_{0}q_{1}q_{2}(1-P_{1/0,1,2}^{d})P_{1/0,1,2}^{0}P_{2/0,1,2}^{d}+\\
+(1-q_{0})q_{1}q_{2}(1-P_{2/1,2}^{d})P_{2/1,2}^{0}(1-P_{1/1,2}^{d})(1-P_{1/1,2}^{0})+q_{0}q_{1}q_{2}(1-P_{2/0,1,2}^{d})P_{2/0,1,2}^{0}(1-P_{1/0,1,2}^{d})(1-P_{1/0,1,2}^{0})+\\
+(1-q_{0})q_{1}q_{2}(1-P_{2/1,2}^{d})P_{2/1,2}^{0}P_{1/1,2}^{d}+q_{0}q_{1}q_{2}(1-P_{2/0,1,2}^{d})P_{2/0,1,2}^{0}P_{1/0,1,2}^{d}\\
\end{aligned}
\end{equation}
}
\begin{equation}
\label{eq:r21}
r_{2}^{1}=(1-q_{0})q_{1}q_{2}(1-P_{1/1,2}^{d})P_{1/1,2}^{0}(1-P_{2/1,2}^{d})P_{2/1,2}^{0}+q_{0}q_{1}q_{2}(1-P_{1/0,1,2}^{d})P_{1/0,1,2}^{0}(1-P_{2/0,1,2}^{d})P_{2/0,1,2}^{0}
\end{equation}

In Fig.~\ref{fig:mc_rxtx} we present the discrete time Markov Chain (DTMC) that describes the queue evolution. Each state is denoted by an integer and represents the queue size at the relay node. The transition matrix of the above DTMC is is a lower Hessenberg matrix given by:
\begin{equation}
P=\left(\begin{array}{ccccc}
a_{0} & b_{0} & 0  & 0    & \cdots   \\
a_{1} & b_{1} & b_{0} & 0 & \cdots    \\
a_{2} & b_{2} & b_{1} & b_{0} & \cdots \\
0 & b_{3} & b_{2} & b_{1} & \cdots \\
0 & 0 & b_{3} & b_{2} & \cdots\\
\vdots & \vdots & \vdots & \vdots & \ddots
\end{array} \right)
\end{equation}

Where $a_{0}=1-p_{1}^{0}-p_{2}^{0},a_{1}=p_{1}^{0},a_{2}=p_{2}^{0}$, $b_{0}=p_{-1}^{1}$ and $b_{i+1}=p_{i}^{1} \text{ }  i=0,1,2,3$. The quantity $p_{i}^{0}$ ($p_{i}^{1}$) is the probability that the queue size increases by $i$ packets when the queue is empty (not empty). Note that $p_{i}^{0}=r_{i}^{0}$, because when the queue is empty the probability of $i$ packets arriving is the same with the probability that the queue size increases by $i$ packets; when the queue is not empty however, this is not true. For example the probability of $2$ packets arriving is not the same with the probability of increasing the queue size by $2$; this is because both arrivals and departures can occur at the same time. The expressions for the $p_{i}^{1}$ are also given by lengthy expressions listed below:

{\scriptsize
\begin{equation}
\label{eq:p-11}
\begin{aligned}
p_{-1}^{1}=q_{0}(1-q_{1})(1-q_{2})P_{0/0}^{d}+q_{0}(1-q_{1})q_{2}P_{0/0,2}^{d}P_{2/0,2}^{d}+q_{0}(1-q_{1})q_{2}P_{0/0,2}^{d}(1-P_{2/0,2}^{d})(1-P_{2/0,2}^{0})+\\
q_{0}q_{1}(1-q_{2})P_{0/0,1}^{d}P_{1/0,1}^{d}+q_{0}q_{1}(1-q_{2})P_{0/0,1}^{d}(1-P_{1/0,1}^{d})(1-P_{1/0,1}^{0})+q_{0}q_{1}q_{2}P_{0/0,1,2}^{d}P_{1/0,1,2}^{d}P_{2/0,1,2}^{d}+\\
+q_{0}q_{1}q_{2}P_{0/0,1,2}^{d}(1-P_{1/0,1,2}^{d})(1-P_{1/0,1,2}^{0})(1-P_{2/0,1,2}^{d})(1-P_{2/0,1,2}^{0})+q_{0}q_{1}q_{2}P_{0/0,1,2}^{d}P_{1/0,1,2}^{d}(1-P_{2/0,1,2}^{d})(1-P_{2/0,1,2}^{0})+\\
q_{0}q_{1}q_{2}P_{0/0,1,2}^{d}(1-P_{1/0,1,2}^{d})(1-P_{1/0,1,2}^{0})P_{2/0,1,2}^{d}
\end{aligned}
\end{equation}

\begin{equation}
\label{eq:p01}
\begin{aligned}
p_{0}^{1}=1-p_{-1}^{1}-p_{1}^{1}-p_{2}^{1}
\end{aligned}
\end{equation}

\begin{equation}
\label{eq:p11}
\begin{aligned}
p_{1}^{1}=(1-q_{0})q_{1}(1-q_{2})(1-P_{1/1}^{d})P_{1/1}^{0}+(1-q_{0})q_{1}q_{2}(1-P_{1/1,2}^{d})P_{1/1,2}^{0}P_{2/1,2}^{d}+\\
(1-q_{0})q_{1}q_{2}(1-P_{1/1,2}^{d})P_{1/1,2}^{0}(1-P_{2/1,2}^{d})(1-P_{2/1,2}^{0})+(1-q_{0})(1-q_{1})q_{2}(1-P_{2/2}^{d})P_{2/2}^{0}+\\
(1-q_{0})q_{1}q_{2}(1-P_{2/1,2}^{d})P_{2/1,2}^{0}P_{1/1,2}^{d}+(1-q_{0})q_{1}q_{2}(1-P_{2/1,2}^{d})P_{2/1,2}^{0}(1-P_{1/1,2}^{d})(1-P_{1/1,2}^{0})+\\
+q_{0}q_{1}q_{2}P_{0/0,1,2}^{d}(1-P_{1/0,1,2}^{d})P_{1/0,1,2}^{0}(1-P_{2/0,1,2}^{d})P_{2/0,1,2}^{0}+q_{0}q_{1}(1-q_{2})(1-P_{0/0,1}^{d})(1-P_{1/0,1}^{d})P_{1/0,1}^{0}+\\
+q_{0}q_{1}q_{2}(1-P_{0/0,1,2}^{d})(1-P_{1/0,1,2}^{d})P_{1/0,1,2}^{0}P_{2/0,1,2}^{d}+q_{0}q_{1}q_{2}(1-P_{0/0,1,2}^{d})(1-P_{1/0,1,2}^{d})P_{1/0,1,2}^{0}(1-P_{2/0,1,2}^{d})(1-P_{2/0,1,2}^{0})+\\
+q_{0}q_{2}(1-q_{1})(1-P_{0/0,2}^{d})(1-P_{2/0,2}^{d})P_{2/0,2}^{0}+q_{0}q_{1}q_{2}(1-P_{0/0,1,2}^{d})(1-P_{2/0,1,2}^{d})P_{2/0,1,2}^{0}P_{1/0,1,2}^{d}+\\
+q_{0}q_{1}q_{2}(1-P_{0/0,1,2}^{d})(1-P_{2/0,1,2}^{d})P_{2/0,1,2}^{0}(1-P_{1/0,1,2}^{d})(1-P_{1/0,1,2}^{0})
\end{aligned}
\end{equation}

\begin{equation}
\label{eq:p21}
\begin{aligned}
p_{2}^{1}=(1-q_{0})q_{1}q_{2}(1-P_{1/1,2}^{d})P_{1/1,2}^{0}(1-P_{2/1,2}^{d})P_{2/1,2}^{0}+q_{0}q_{1}q_{2}(1-P_{0/0,1,2}^{d})(1-P_{1/0,1,2}^{d})P_{1/0,1,2}^{0}(1-P_{2/0,1,2}^{d})P_{2/0,1,2}^{0}
\end{aligned}
\end{equation}
}

The difference equations that govern the evolution of the states are given by:
\begin{equation}
Ps=s \Rightarrow s_{i}=a_{i}s_{0}+\sum_{j=1}^{i+1}{b_{i-j+1}s_{j}}
\end{equation}

We apply the Z-transform technique to compute the steady state distribution, i.e. we let
\begin{equation}
A(z)=\sum_{i=0}^{2}{a_{i}z^{-i}}, B(z)=\sum_{i=0}^{3}{b_{i}z^{-i}}, S(z)=\sum_{i=0}^{\infty}{s_{i}z^{-i}}
\end{equation}
It is known that~\cite{b:Gebali}:
\begin{equation}
S(z)=s_{0}\frac{z^{-1}A(z)-B(z)}{z^{-1}-B(z)}
\end{equation}

It is also known that the probability of the queue in the relay is empty is given by~\cite{b:Gebali}:
\begin{equation}
\label{eq:P(Q=0)}
P\left(Q=0\right)=\frac{1+B^{'}(1)}{1+B^{'}(1)-A^{'}(1)}
\end{equation}
The expressions of $A^{'}(1)$ and $B^{'}(1)$ are:

\begin{equation}
\label{eq:A'(1)}
\begin{aligned}
A^{'}(z)=\left(\sum_{i=0}^{2}{a_{i}z^{-i}} \right)^{'}=-\sum_{i=1}^{2}{ia_{i}z^{-(i+1)}}
\Rightarrow A^{'}(1)=-\sum_{i=1}^{2}{ia_{i}}\Rightarrow A^{'}(1)=-\sum_{i=1}^{2}{ip_{i}^{0}}=-\lambda_{0}
\end{aligned}
\end{equation}

\begin{equation}
\label{eq:B'(1)}
\begin{aligned}
B^{'}(z)=\left(\sum_{i=0}^{3}{b_{i}z^{-i}} \right)^{'}=-\sum_{i=0}^{3}{ib_{i}z^{-(i+1)}}\Rightarrow B^{'}(1)=-\sum_{i=0}^{3}{ib_{i}}=-b_{1}- 2b_{2}-3b_{3} =-1+p_{-1}^{1}-p_{1}^{1}-2p_{2}^{1}
\end{aligned}
\end{equation}

Then the the probability of the queue in the relay is empty is
\begin{equation}
\label{eq:probempty2}
P\left(Q=0\right)=\frac{p_{-1}^{1}-p_{1}^{1}-2p_{2}^{1}}{p_{-1}^{1}-p_{1}^{1}-2p_{2}^{1}+\lambda_{0}}
\end{equation}

So, the average arrival rate $\lambda$ is given by:
\begin{equation}
\lambda=\frac{p_{-1}^{1}-p_{1}^{1}-2p_{2}^{1}}{p_{-1}^{1}-p_{1}^{1}-2p_{2}^{1}+\lambda_{0}}\lambda_{0}+\frac{\lambda_{0}}{p_{-1}^{1}-p_{1}^{1}-2p_{2}^{1}+\lambda_{0}}\lambda_{1}
\end{equation}

\subsubsection{Condition for the stability of the queue}
An important tool to determine stability is Loyne's criterion~\cite{b:Loynes}, which states that if the arrival and service processes of a queue are jointly strictly stationary and ergodic, the queue is stable if and only if the average arrival rate is strictly less than the average service rate. If the queue is stable, the departure rate (throughput) is equal to the arrival rate.
$\lambda_{1}<\mu\Leftrightarrow r_{1}^{1}+2r_{2}^{1}<\mu$ where $r_{1}^{1}=(1-q_{0})A_{1}+q_{0}B_{1}$, $r_{2}^{1}=(1-q_{0})A_{2}+q_{0}B_{2}$ and $\mu=q_{0}A$.
The expressions for $A,A_{i},B_{i}$ are given by:

{\scriptsize
\begin{equation}
\label{eq:A1B1}
\begin{aligned}
A_{1}=q_{1}(1-q_{2})(1-P_{1/1}^{d})P_{1/1}^{0}+q_{2}(1-q_{1})(1-P_{2/2}^{d})P_{2/2}^{0}+q_{1}q_{2}(1-P_{1/1,2}^{d})P_{1/1,2}^{0}(1-P_{2/1,2}^{d})(1-P_{2/1,2}^{0})+\\
+q_{1}q_{2}(1-P_{1/1,2}^{d})P_{1/1,2}^{0}P_{2/1,2}^{d}+q_{1}q_{2}(1-P_{2/1,2}^{d})P_{2/1,2}^{0}(1-P_{1/1,2}^{d})(1-P_{1/1,2}^{0})+q_{1}q_{2}(1-P_{2/1,2}^{d})P_{2/1,2}^{0}P_{1/1,2}^{d}\\
\\
B_{1}=q_{1}(1-q_{2})(1-P_{1/0,1}^{d})P_{1/0,1}^{0}+q_{2}(1-q_{1})(1-P_{2/0,2}^{d})P_{2/0,2}^{0}+q_{1}q_{2}(1-P_{1/0,1,2}^{d})P_{1/0,1,2}^{0}(1-P_{2/0,1,2}^{d})(1-P_{2/0,1,2}^{0})+\\
+q_{1}q_{2}(1-P_{1/0,1,2}^{d})P_{1/0,1,2}^{0}P_{2/0,1,2}^{d}+q_{1}q_{2}(1-P_{2/0,1,2}^{d})P_{2/0,1,2}^{0}(1-P_{1/0,1,2}^{d})(1-P_{1/0,1,2}^{0})+\\
+q_{1}q_{2}(1-P_{2/0,1,2}^{d})P_{2/0,1,2}^{0}P_{1/0,1,2}^{d}
\end{aligned}
\end{equation}
}

\begin{equation}
\label{eq:A2B2}
\begin{aligned}
A_{2}=q_{1}q_{2}(1-P_{1/1,2}^{d})P_{1/1,2}^{0}(1-P_{2/1,2}^{d})P_{2/1,2}^{0},\\ B_{2}=q_{1}q_{2}(1-P_{1/0,1,2}^{d})P_{1/0,1,2}^{0}(1-P_{2/0,1,2}^{d})P_{2/0,1,2}^{0}
\end{aligned}
\end{equation}

\begin{equation}
\label{eq:A}
\begin{aligned}
A=(1-q_{1})(1-q_{2})P_{0/0}^{d}+q_{1}(1-q_{2})P_{0/0,1}^{d}+q_{2}(1-q_{1})P_{0/0,2}^{d}+q_{1}q_{2}P_{0/0,1,2}^{d}
\end{aligned}
\end{equation}

Then the values of $q_{0}$ for which the queue is stable is given by $q_{0min}<q_{0}<1$, where:
\begin{equation}
\label{eq:q0min2}
q_{0min}=\frac{A_{1}+2A_{2}}{A+A_{1}+2A_{2}-B_{1}-2B_{2}}
\end{equation}

\subsubsection{Average queue size}
\label{sec:avq}
The average queue size is given by~\cite{b:Gebali}: $\overline{Q}=-S^{'}(1)$ where $S^{'}(1)=s_{0}\frac{K^{''}(1)}{L^{''}(1)}$.
The expressions for $K(z)$ and $L(z)$ are given by:

\begin{equation}
\label{eq:K(z)}
\begin{aligned}
K(z)=\left(-z^{-2}A(z)+z^{-1}A^{'}(z)-B^{'}(z) \right) \left(z^{-1}-B(z)\right) - \left(z^{-1}A(z)-B(z) \right) \left(-z^{-2}-B^{'}(z) \right)
\end{aligned}
\end{equation}

\begin{equation}
\label{eq:L(z)}
\begin{aligned}
L(z)=\left(z^{-1}-B(z) \right)^{2}
\end{aligned}
\end{equation}

Then $K^{''}(1)$ and $L^{''}(1)$ are given by:
\begin{equation}
\label{eq:K''(1)}
\begin{aligned}
\Rightarrow K^{''}(1)=\left(2A(1)-2A^{'}(1)+A^{''}(1)-B^{''}(1) \right) \left(-1-B^{'}(1) \right)- \left(2-B^{''}(1) \right) \left(-A(1)+A^{'}(1)-B^{'}(1) \right)
\end{aligned}
\end{equation}

\begin{equation}
\label{eq:L''(1)}
\begin{aligned}
L^{''}(z)=\left[2\left(z^{-1}-B(z) \right) \left(-z^{-2}-B^{'}(z) \right) \right]^{'} \Rightarrow L^{''}(1)=2\left(-1-B^{'}(1)\right)^{2}
\end{aligned}
\end{equation}

The values of $A^{''}(1)$ and $B^{''}(1)$ are:

\begin{equation}
\label{eq:A''(1)}
\begin{aligned}
A^{''}(z)=\left(-\sum_{i=1}^{2}{ia_{i}z^{-(i+1)}} \right)^{'}=\sum_{i=1}^{2}{i(i+1)a_{i}z^{-(i+2)}}\Rightarrow A^{''}(1)=2p_{1}^{0}+6p_{2}^{0}
\end{aligned}
\end{equation}

\begin{equation}
\label{eq:B''(1)}
\begin{aligned}
B^{''}(z)=\left(-\sum_{i=1}^{3}{ib_{i}z^{-(i+1)}} \right)^{'}=\sum_{i=1}^{3}{i(i+1)b_{i}z^{-(i+2)}}\Rightarrow B^{''}(1)=2-2p_{-1}^{1}+4p_{1}^{1}+10p_{2}^{1}
\end{aligned}
\end{equation}

The average queue size is given by:

\begin{equation}
\label{eq:avQ2}
\overline{Q}=\frac{(p_{1}^{1}+2p_{2}^{1}-p_{-1}^{1})(4p_{1}^{0}+10p_{2}^{0})+\lambda_{0}(2p_{-1}^{1}-4p_{1}^{1}-10p_{2}^{1})}{2(p_{1}^{1}+2p_{2}^{1}-p_{-1}^{1})(p_{-1}^{1}-p_{1}^{1}-2p_{2}^{1}+\lambda_{0})}
\end{equation}

\subsubsection{The throughput per user and the aggregate throughput}
The throughput rates $\mu_{1}, \mu_{2}$ for the users $1,2$ are given by:

\begin{equation}
\label{eq:thr2_1}
\begin{aligned}
\mu_{1}=q_{0}P\left(Q>0\right)q_{1}(1-q_{2})\left[P_{1/0,1}^{d}+(1-P_{1/0,1}^{d})P_{1/0,1}^{0}\right]+q_{0}P\left(Q>0\right)q_{1}q_{2}\left[P_{1/0,1,2}^{d}+(1-P_{1/0,1,2}^{d})P_{1/0,1,2}^{0} \right]+\\
+\left[1-q_{0}P\left(Q>0\right)\right]q_{1}(1-q_{2})\left[P_{1/1}^{d}+(1-P_{1/1}^{d})P_{1/1}^{0} \right]+\left[1-q_{0}P\left(Q>0\right)\right]q_{1}q_{2}\left[P_{1/1,2}^{d}+(1-P_{1/1,2}^{d})P_{1/1,2}^{0} \right]
\end{aligned}
\end{equation}

\begin{equation}
\label{eq:thr2_2}
\begin{aligned}
\mu_{2}=q_{0}P\left(Q>0\right)q_{2}(1-q_{1})\left[P_{2/0,2}^{d}+(1-P_{2/0,2}^{d})P_{2/0,2}^{0}\right]+q_{0}P\left(Q>0\right)q_{1}q_{2}\left[P_{2/0,1,2}^{d}+(1-P_{2/0,1,2}^{d})P_{2/0,1,2}^{0} \right]+\\
+\left[1-q_{0}P\left(Q>0\right)\right]q_{2}(1-q_{1})\left[P_{2/2}^{d}+(1-P_{2/2}^{d})P_{2/2}^{0} \right]+\left[1-q_{0}P\left(Q>0\right)\right]q_{1}q_{2}\left[P_{2/1,2}^{d}+(1-P_{2/1,2}^{d})P_{2/1,2}^{0} \right]
\end{aligned}
\end{equation}

In the equations above we assume that the queue is stable, hence the arrival rate from each user to the queue is a contribution to its overall throughput. The aggregate throughput is $\mu_{total}=\mu_{1}+\mu_{2}$. Notice that the throughput per user is independent of $q_{0}$ as long as it is in the stability region. This is explained because the product $q_{0}P\left(Q>0\right)$ is constant. The proof is straightforward and thus is omitted.

When the queue is unstable however, the aggregate throughput is the summation of all the direct throughput between the users and the destination plus the service rate of the relay.

\subsection{N-symmetric users}
We now generalize the above for the case of a symmetric $n$-users network. Each user attempts to transmit in a slot with probability $q$; the success probability to the relay and the destination when $i$ nodes transmit are given by $P_{0,i}$, $P_{d,i}$  respectively. There are two cases for the $P_{d,i}$ , $P_{d,i,0}$, $P_{d,i,1}$  denoting success probability when relay remains silent or transmits respectively. The above success probabilities for the symmetric case are given by $P_{d,i,j}=P_{d}\left(\frac{1}{1+\gamma_{d}} \right)^{i-1} \left(\frac{1}{1+\beta\gamma_{0}} \right)^{j},\text{ } j=0,1$ and $\beta=\frac{v_{0d}g_{0d}}{v_{d}g_{d}}>1$. $P_{0d,i}=P_{0d}\left(\frac{1}{1+\frac{1}{\beta}\gamma_{d}}\right)^{i}$, $P_{0}=\exp\left(-\frac{\gamma_{0}\eta_{0}}{v_{0}g_{0}}\right)$, $P_{d}=\exp\left(-\frac{\gamma_{d}\eta_{d}}{v_{d}g_{d}}\right)$, $P_{0d}=\exp\left(-\frac{\gamma_{0}\eta_{0}}{v_{0}g_{0}}\right)$. There are two cases for the $P_{0,i}$, $P_{0,i,0},P_{0,i,1}$ denoting success probability when relay remains silent or transmits respectively. The success probabilities are given by $P_{0,i,0}=P_{0}\left(\frac{1}{1+\gamma_{0}} \right)^{i-1}$ and
$P_{0,i,1}=P_{0}\left(1+\gamma_{0}r_{0}^{\alpha}g \right)^{-1}\left(\frac{1}{1+\gamma_{0}} \right)^{i-1}$ where $r_{0}$ is the distance between the users and the relay, $\alpha$ is the path loss exponent and $g$ is the self interference coefficient.

\subsubsection{Computation of the average arrival and service rate}
The service rate is given by the following equation:
\begin{equation}
\label{eq:mun}
\mu=\sum_{k=0}^{n}{{n \choose k} {q_{0}q^{k}(1-q)^{n-k}}P_{0d,k}}
\end{equation}

The average arrival rate $\lambda$ of the queue is given by:
\begin{equation}
\label{eq:lamdan}
\lambda=P\left(Q=0\right)\lambda_{0}+P\left(Q>0\right)\lambda_{1}
\end{equation}

$\lambda_{0}=\sum_{k=1}^{n}{kr_{k}^{0}}$ where the $r_{k}^{0}$ is the probability that the relay received $k$ packets when the queue is empty, the expression for $r_{k}^{0}$ is given by:

\begin{equation}
\label{eq:rk0n}
\begin{aligned}
r_{k}^{0}=\sum_{i=k}^{n}{{n \choose i}{i \choose k} {q^{i}(1-q)^{n-i}}P_{0,i,0}^{k}\left(1-P_{d,i,0}\right)^{k}\left[1-P_{0,i,0}(1-P_{d,i,0})\right]^{i-k}},\text{ }1 \leq k \leq n
\end{aligned}
\end{equation}

$\lambda_{1}=\sum_{k=1}^{n}{kr_{k}^{1}}$ where the $r_{k}^{1}$ is the probability that the relay received $k$ packets when the queue is not empty and is given by:
\begin{equation}
\label{eq:rk1n}
\begin{aligned}
r_{k}^{1}=(1-q_{0})\sum_{i=k}^{n}{{n \choose i}{i \choose k} {q^{i}(1-q)^{n-i}}P_{0,i,0}^{k}\left(1-P_{d,i,0}\right)^{k}\left[1-P_{0,i,0}(1-P_{d,i,0})\right]^{i-k}}+\\
+q_{0}\sum_{i=k}^{n}{{n \choose i}{i \choose k} {q^{i}(1-q)^{n-i}}P_{0,i,1}^{k}\left(1-P_{d,i,1}\right)^{k}\left[1-P_{0,i,1}(1-P_{d,i,1})\right]^{i-k}},\text{ }1 \leq k \leq n
\end{aligned}
\end{equation}

The elements of the transition matrix are given by:
$a_{k}=p_{k}^{0}$, $b_{0}=p_{-1}^{1}$, $b_{1}=p_{0}^{1}$ and $b_{k+1}=p_{k}^{1} \text{ } \forall k>0$ where:

\begin{equation}
\label{eq:pk0n}
p_{k}^{0}=\sum_{i=k}^{n}{{n \choose i}{i \choose k} {q^{i}(1-q)^{n-i}}P_{0,i,0}^{k}\left(1-P_{d,i,0}\right)^{k}\left[1-P_{0,i,0}(1-P_{d,i,0})\right]^{i-k}},\text{ }1 \leq k \leq n
\end{equation}

\begin{equation}
\label{eq:p-11n}
p_{-1}^{1}=q_{0}\sum_{k=0}^{n}{{n \choose k}q^{k}(1-q)^{n-k}P_{0d,k}\left[1-P_{0,k,1}(1-P_{d,k,1})\right]^{k}}
\end{equation}

\begin{equation}
\label{eq:pk1n}
\begin{aligned}
p_{k}^{1}=(1-q_{0})\sum_{i=k}^{n}{{n \choose i}{i \choose k} {q^{i}(1-q)^{n-i}}P_{0,i,0}^{k}\left(1-P_{d,i,0}\right)^{k}\left[1-P_{0,i,0}(1-P_{d,i,0})\right]^{i-k}}+\\
+q_{0}\sum_{i=k}^{n}{{n \choose i}{i \choose k} {q^{i}(1-q)^{n-i}}(1-P_{0d,i})P_{0,i,1}^{k}\left(1-P_{d,i,1}\right)^{k}\left[1-P_{0,i,1}(1-P_{d,i,1})\right]^{i-k}}+\\
+q_{0}\sum_{i=k+1}^{n}{{n \choose i}{i \choose k+1} {q^{i}(1-q)^{n-i}}P_{0d,i}P_{0,i,1}^{k+1}\left(1-P_{d,i,1}\right)^{k+1}\left[1-P_{0,i,1}(1-P_{d,i,1})\right]^{i-k-1}}
\end{aligned}
\end{equation}

\begin{equation}
\label{eq:p01n}
p_{0}^{1}=1-p_{-1}^{1}-\sum_{i=1}^{n}{p_{i}^{1}}
\end{equation}

The probability that the queue in the relay is empty is given by (\ref{eq:P(Q=0)}), the expressions for $A^{'}(1)$ and $B^{'}(1)$ are:

\begin{equation}
\label{eq:A'(1)}
\begin{aligned}
A^{'}(z)=\left(\sum_{i=0}^{n}{a_{i}z^{-i}} \right)^{'}=-\sum_{i=1}^{n}{ia_{i}z^{-(i+1)}}
\Rightarrow A^{'}(1)=-\sum_{i=1}^{n}{ia_{i}}\Rightarrow A^{'}(1)=-\sum_{i=1}^{n}{ip_{i}^{0}}=-\lambda_{0}
\end{aligned}
\end{equation}

\begin{equation}
\label{eq:B'(1)}
\begin{aligned}
B^{'}(z)=\left(\sum_{i=0}^{n+1}{b_{i}z^{-i}} \right)^{'}=-\sum_{i=i}^{n+1}{ib_{i}z^{-(i+1)}}\Rightarrow B^{'}(1)=-\sum_{i=i}^{n+1}{ib_{i}}=-b_{1}-\sum_{i=2}^{n+1}{ib_{i}}=-1+p_{-1}^{1}-\sum_{i=1}^{n}{ip_{i}^{1}}
\end{aligned}
\end{equation}

Then the probability that the queue in the relay is empty is given by:

\begin{equation}
P\left( Q=0 \right)=\frac{\displaystyle p_{-1}^{1}-\sum_{i=1e}^{n}{ip_{i}^{1}}}{\displaystyle p_{-1}^{1}-\sum_{i=1}^{n}{ip_{i}^{1}}+\lambda_{0}}
\end{equation}

\subsubsection{Condition for the stability of the queue}
$\lambda_{1}<\mu\Leftrightarrow \sum_{k=1}^{n}{kr_{k}^{1}}<\mu$ where $r_{k}^{1}=(1-q_{0})A_{k}+q_{0}B_{k}$ and $\mu=q_{0}A$.
The expressions for $A,A_{k},B_{k}$ are :
\begin{equation}
\label{eq:Akn}
\begin{aligned}
A_{k}=\sum_{i=k}^{n}{{n \choose i}{i \choose k} {q^{i}(1-q)^{n-i}}P_{0,i,0}^{k}\left(1-P_{d,i,0}\right)^{k}\left[1-P_{0,i,0}(1-P_{d,i,0})\right]^{i-k}}
\end{aligned}
\end{equation}

\begin{equation}
\label{eq:Bkn}
\begin{aligned}
B_{k}=\sum_{i=k}^{n}{{n \choose i}{i \choose k} {q^{i}(1-q)^{n-i}}P_{0,i,1}^{k}\left(1-P_{d,i,1}\right)^{k}\left[1-P_{0,i,1}(1-P_{d,i,1})\right]^{i-k}}
\end{aligned}
\end{equation}

\begin{equation}
\label{eq:An}
\begin{aligned}
A=\sum_{k=0}^{n}{{n \choose k} {q^{k}(1-q)^{n-k}}P_{0d,k}}
\end{aligned}
\end{equation}

The values of $q_{0}$ for which the queue is stable is given by $q_{0min}<q_{0}<1$, where:
\begin{equation}
\label{eq:q0minn}
q_{0min}=\frac{\displaystyle \sum_{k=1}^{n}{kA_{k}}}{\displaystyle A+\sum_{k=1}^{n}{kA_{k}}-\sum_{k=1}^{n}{kB_{k}}}
\end{equation}

\subsubsection{Average queue size}
As we showed in the Section~\ref{sec:avq}, the average queue size is given by: $\overline{Q}=-S^{'}(1)$ where $S^{'}(1)=s_{0}\frac{K^{''}(1)}{L^{''}(1)}$.
The expressions for $K^{''}(1)$ and $L^{''}(1)$ are given by (\ref{eq:K''(1)}) and (\ref{eq:L''(1)}).
The expressions for $A^{''}(1)$ and $B^{''}(1)$ are:
\begin{equation}
\label{eq:A''(1)}
\begin{aligned}
A^{''}(z)=\left(-\sum_{i=1}^{n}{ia_{i}z^{-(i+1)}} \right)^{'}=\sum_{i=1}^{n}{i(i+1)a_{i}z^{-(i+2)}}\Rightarrow A^{''}(1)=\sum_{i=1}^{n}{i(i+1)a_{i}}=\sum_{i=1}^{n}{i(i+1)p_{i}^{0}}
\end{aligned}
\end{equation}

\begin{equation}
\label{eq:B''(1)}
\begin{aligned}
B^{''}(z)=\left(-\sum_{i=i}^{n+1}{ib_{i}z^{-(i+1)}} \right)^{'}=\sum_{i=1}^{n+1}{i(i+1)b_{i}z^{-(i+2)}}\Rightarrow B^{''}(1)=\sum_{i=1}^{n+1}{i(i+1)b_{i}}=2-2p_{-1}^{1}+\sum_{i=1}^{n}{i(i+3)p_{i}^{1}}
\end{aligned}
\end{equation}

Following the same methodology as in Section~\ref{sec:avq} we obtain that the average queue size is given by:

\begin{equation}
\label{eq:avQn}
\begin{aligned}
\overline{Q}=\frac{\displaystyle \left(\sum_{i=1}^{n}{ip_{i}^{1}}-p_{-1}^{1} \right)\sum_{i=1}^{n}{i(i+3)p_{i}^{0}}+\lambda_{0}\left(2p_{-1}^{1}-\sum_{i=1}^{n}{i(i+3)p_{i}^{1}} \right)}{\displaystyle 2\left(\sum_{i=1}^{n}{ip_{i}^{1}}-p_{-1}^{1} \right) \left(p_{-1}^{1}-\sum_{i=1}^{n}{ip_{i}^{1}}+\lambda_{0} \right)}
\end{aligned}
\end{equation}

\subsubsection{The throughput per user and the aggregate throughput}
The throughput per user for the network with the relay when the queue is stable is given by:

\begin{equation}
\label{eq:thrn}
\begin{aligned}
\mu=q_{0}P\left(Q>0\right)\sum_{k=0}^{n-1}{{n-1 \choose k}q^{k+1}(1-q)^{n-1-k}\left[P_{d,k+1,1}+(1-P_{d,k+1,1})P_{0,k+1,1}\right]}+ \\
+\left[1-q_{0}P\left(Q>0\right)\right]\sum_{k=0}^{n-1}{{n-1 \choose k}q^{k+1}(1-q)^{n-1-k}\left[P_{d,k+1,0}+\left(1-P_{d,k+1,0}\right)P_{0,k+1,0}\right]}
\end{aligned}
\end{equation}

When the queue is unstable though, the throughput per user is given by the summation of the direct to the destination throughput plus the service rate of the relay divided by the number of the users $n$.
The aggregate throughput is $\mu_{total}=n\mu$.

\section{Numerical Results}
\label{sec:results}
In this section we present numerical results for the analysis presented above. To simplify the presentation we consider the case where all the users have the same link characteristics and transmission probabilities. The parameters used in the numerical results are as follows. The distances in meters are given by $r_{d}=130$, $r_{0}=60$ and $r_{0d}=80$. The path loss is $\alpha=4$ and the receiver noise power $\eta=10^{-11}$. The transmit power for the relay is $P_{tx}(0)=10$ mW and for the i-th user $P_{tx}(i)=1$ mW. We used $\gamma<1$ because it is possible for two or more users to transmit successfully at the same time.

\begin{figure}[h!]
\centering
\subfigure[Throughput per user vs self interference coefficient]{
\includegraphics[scale=0.45]{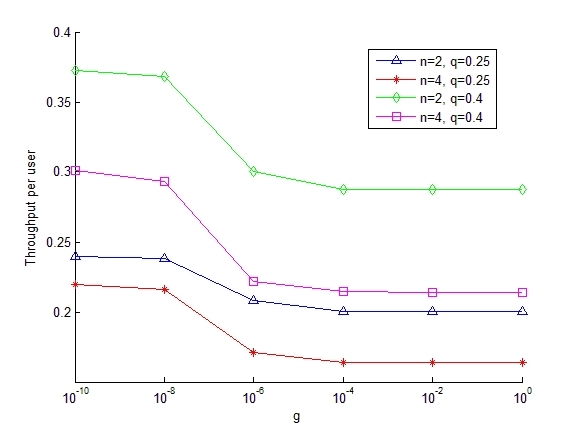}
\label{fig:thr_02}
}
\subfigure[Aggregate throughput vs self interference coefficient]{
\includegraphics[scale=0.45]{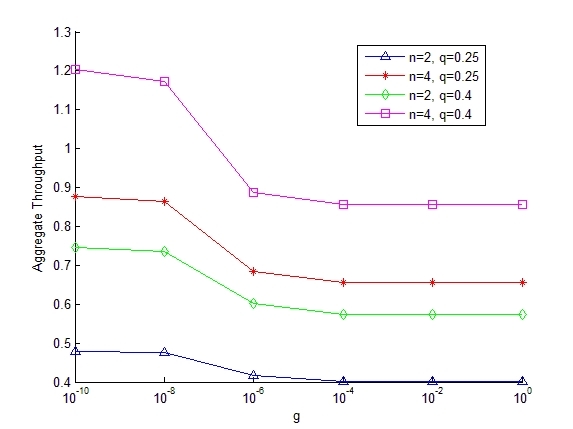}
\label{fig:athr_02}
}
\label{fig:g02}
\caption{Throughput per user and aggregate throughput vs the self interference coefficient for $\gamma=0.2$}
\end{figure}

\begin{figure}[h!]
\centering
\subfigure[Throughput per user vs self interference coefficient]{
\includegraphics[scale=0.45]{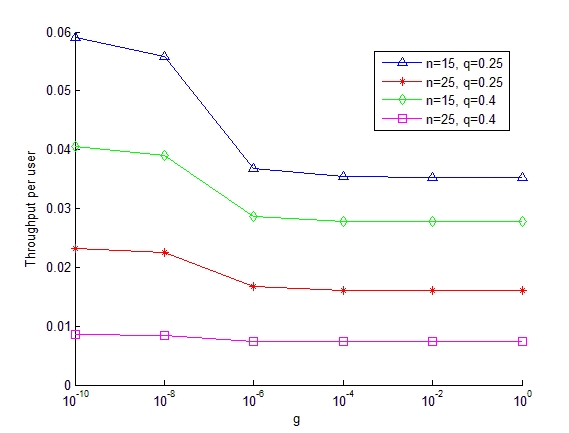}
\label{fig:thr_06}
}
\subfigure[Aggregate throughput vs self interference coefficient]{
\includegraphics[scale=0.45]{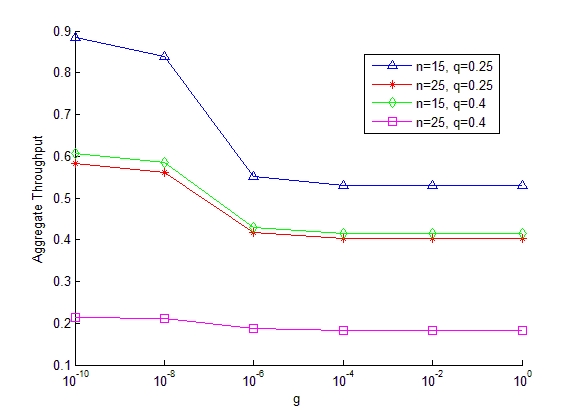}
\label{fig:athr_06}
}
\label{fig:g06}
\caption{Throughput per user and aggregate throughput vs the self interference coefficient for $\gamma=0.6$}
\end{figure}

The figures~\ref{fig:thr_02} and ~\ref{fig:thr_06} present the throughput per user versus $g$ (the self-interference coefficient) for various values of $q$,$\gamma$ and $n$. The figures~\ref{fig:athr_02} and~\ref{fig:athr_06} show the aggregate throughput versus $g$.

\begin{figure}[h!]
\centering
\subfigure[Throughput per user vs the number of the users]{
\includegraphics[scale=0.5]{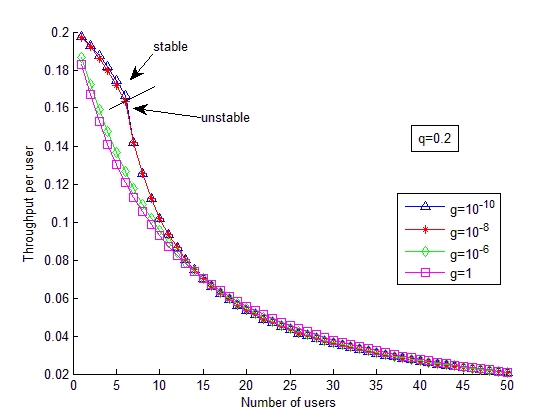}
\label{fig:thr_n_02}
}
\subfigure[Aggregate throughput vs the number of the users]{
\includegraphics[scale=0.5]{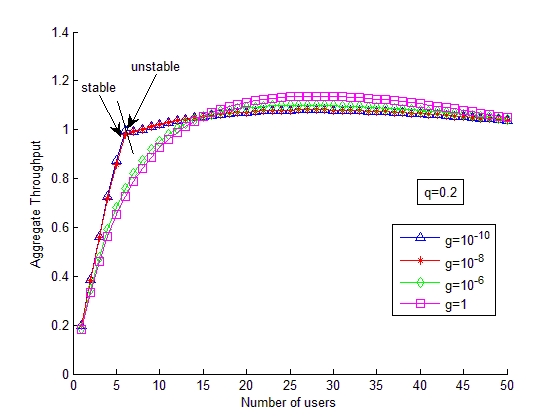}
\label{fig:athr_n_02}
}
\label{fig:gn02}
\caption{Throughput per user and aggregate throughput vs the number of the users for $\gamma=0.2$}
\end{figure}

\begin{figure}[h!]
\centering
\subfigure[Throughput per user vs the number of the users]{
\includegraphics[scale=0.5]{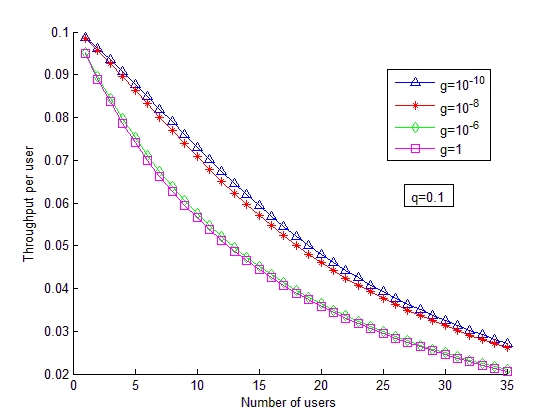}
\label{fig:thr_n_06}
}
\subfigure[Aggregate throughput vs the number of the users]{
\includegraphics[scale=0.5]{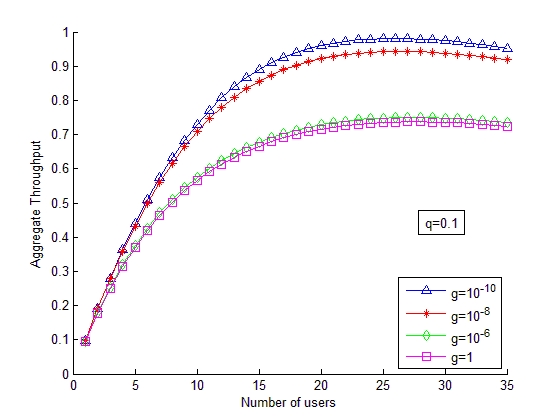}
\label{fig:athr_n_06}
}
\label{fig:gn06}
\caption{Throughput per user and aggregate throughput vs the number of the users for $\gamma=0.6$}
\end{figure}

The figures~\ref{fig:thr_n_02} and ~\ref{fig:thr_n_06} present the throughput per user versus the number of the users in the network for various values of $q$,and $g$ for $\gamma=0.2$ and $\gamma=0.6$ respectively. The figures~\ref{fig:athr_n_02} and~\ref{fig:athr_n_06} show the aggregate throughput versus the number of the users.
When $\gamma=0.2$ we observe that for $g=10^{-10}$ and $g=10^{-8}$ (almost perfect self-interference cancelation) the relay's queue is unstable for relative small number of users. The previous result is because the small value of $\gamma$ is more likely more transmissions from the users to the relay to be successful, but at the same time the relay can transmit at most one packet per time slot. For $\gamma=0.6$ the queue is never unstable for the parameters described in the figures, and for $g=10^{-10}$ and $g=10^{-8}$ the advantages in term of throughput are obvious compared to no self interference cancelation.

\begin{figure}[h!]
\centering
\subfigure[Stability threshold vs the number of the users for $\gamma=0.2$]{
\includegraphics[scale=0.45]{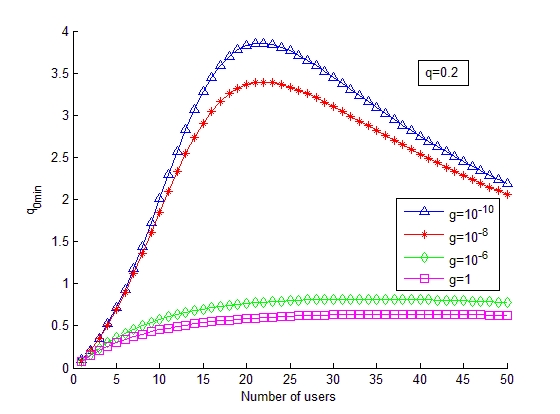}
\label{fig:q0min_n_02}
}
\subfigure[Stability threshold vs the number of the users for $\gamma=0.6$]{
\includegraphics[scale=0.45]{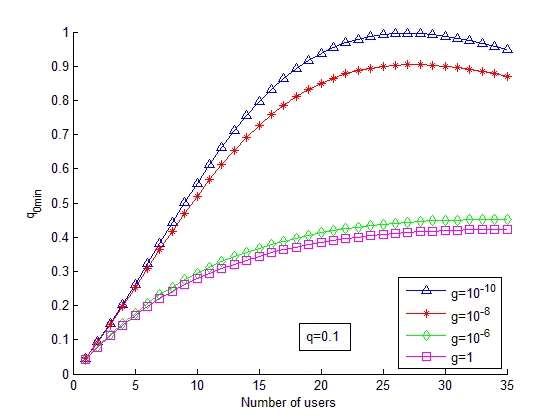}
\label{fig:q0min_n_06}
}
\label{fig:q0min}
\caption{Stability threshold vs the number of the users}
\end{figure}

The Fig.~\ref{fig:q0min_n_02} and Fig.~\ref{fig:q0min_n_06} show the $q_{0min}$ vs $n$ for $\gamma=0.2$ and $\gamma=0.6$ respectively. Note that $q_{0min}<q<1$, so when $q_{0min}\geq 1$ then the queue is unstable as in case of $\gamma=0.2$ for $g=10^{-10}$ and $g=10^{-8}$.

\section{Conclusions}
\label{sec:conclusions}
In this paper, we examined the operation of a node relaying packets from a number of users to a common destination node.
We assumed MPR capability for the relay and for the destination node. We studied a multiple capture model, where a user's transmission is successful if the received $SINR$ is above a threshold $\gamma$. The relay node can also receive and transmit simultaneously, so the problem of self interference arises.

We obtained analytical expressions for the relay's queue characteristics such as the stability condition, the values of the arrival and service rates, the average queue size. We studied the throughput per user and the aggregate throughput, and found that, under stability conditions, the throughput per user does not depend on the relay probability of transmission. We studied the impact of self interference coefficient on the throughput per user and the aggregate throughput of the network.

We showed that for perfect self-interference cancelation, the advantages are obvious. Another interesting result is that the self interference coefficient plays a crucial role when $\gamma$ is small (and $g$ tends to zero) because it can easily cause an unstable queue.

Future extensions of this work should include users with non-saturated queues i.e. sources with external random arrivals, a relay node with its own packets and different priorities for the users.

{\footnotesize
\bibliographystyle{unsrt}
\bibliography{bibliography}}

\end{document}